\documentclass[12pt]{article}
\usepackage{latexsym}
\usepackage{epsf}
\makeatletter
\newcommand{\vereq}[2]{\lower3pt\vbox{\baselineskip1.5pt \lineskip1.5pt
\ialign{$\m@th#1\hfill##\hfil$\crcr#2\crcr\sim\crcr}}}
\makeatother
\newcommand{\lessim}{\mathrel{\mathpalette\vereq<}}

\newcommand{\pf }{{\rm Pf}}
\newcommand{\lb }{\left[}
\newcommand{\rb }{\right]}
\newcommand{\lp }{\left(}
\newcommand{\rp }{\right)}

\newcommand{\CR}{{\cal R}}

\newcommand{\Dsl}{\hbox{{\ {/\kern-.6600em $D$}}}}
\newcommand{\ssDsl}{\hbox{{\ {${\scriptstyle /}$\kern-.6250em ${\scriptstyle D}$
}}}}
\newcommand{\dsl}{\hbox{{\ {/\kern-.5500em $\partial$}}}}

\newcommand{\eqn}[1]{\label{eq:#1}}

\newcommand{\beq}{\begin{eqnarray}}
\newcommand{\eeq}{\end{eqnarray}}
\begin{document}
\title{Supersymmetric Yang-Mills Theories from\\ Domain Wall
  Fermions\footnote{Talk given by D.K. at  CHIRAL '99, Taipei,
    Sep. 13-18, 1999.} }
\author{David B. Kaplan $^a$ and Martin Schmaltz $^b$\\
           \vbox{\vskip 1.5truecm}
       \normalsize \it  $^a$Institute for Nuclear Theory, University of Washington\\
           \normalsize \it  Seattle, WA 98195-1550, USA \\
          { \normalsize\tt dbkaplan@phys.washington.edu} \\
           \vbox{\vskip 1.0truecm}
           \normalsize  \it   $^b$SLAC, Stanford University, Stanford, CA 94309, USA \\
          { \normalsize\tt schmaltz@slac.stanford.edu}\\
           \vbox{\vskip 0.5truecm} }
\vspace{2cm}
\date{ }
\maketitle
\begin{abstract}
%
We present work in progress on employing domain wall
fermions to simulate $N=1$ supersymmetric
Yang-Mills theories on the lattice in $d=4$ and $d=3$ dimensions.  The 
geometrical nature of domain wall fermions gives simple insights into
how to construct these theories.  We
also discuss the obstacles associated with simulating the $N=2$ theory 
in $d=4$. 
\end{abstract}
%
%
\section{Chirality and accidental supersymmetry}

There has been intense interest in supersymmetry (SUSY) in the past two
decades. The past several years have witnessed many interesting and
compelling speculations about strongly coupled SUSY
theories.  It would be interesting to test these conjectures on the
lattice.  However, since Poincar\'e symmetry does not exist on the
lattice, supersymmetry does not either. In principle, one could tune
lattice theories to the SUSY critical point. However, just as
Poincar\'e symmetry is recovered without fine tuning, one might hope
that SUSY could be similarly obtained in the continuum limit.

The secret to why the Poincar\'e symmetric point takes no work to find 
is that the imposition of hypercubic symmetry and gauge symmetry
ensures that Poincar\'e symmetry is an {\it accidental} symmetry. 
All allowed operators that violate the symmetry are
irrelevant. Therefore the continuum limit of the theory automatically
exhibits more symmetry than it possesses at finite lattice spacing. If
supersymmetry could arise as an accidental
symmetry as well, then simulation of such theories would not entail
fine-tuning of parameters, and would be relatively simple.

The outlook for this approach in SUSY theories with scalar
fields is poor...scalar mass terms violate SUSY, are
relevant, and cannot be forbidden by any symmetry, unless the scalars
are Goldstone bosons.  I
will return to this issue later, when we talk about $N=2$ super
Yang-Mills (SYM) theories. 

However, there are some
SUSY theories which do not entail scalars.  Of particular interest is
$N=1$ SYM theory
in $d=4$  dimensions. In this theory, the only relevant SUSY violating parameter is the
gaugino mass, which can be forbidden by a discrete chiral
symmetry. Thus if one can realize the chiral symmetry  on the lattice, $N=1$ SUSY
can arise as an accidental symmetry in the continuum limit.  This is
where domain wall fermions \cite{Kaplan:1993sg,Shamir:1993zy} come in,
for which  chiral  
symmetry violation (for  weak coupling) tends to zero exponentially fast 
in the  domain wall separation.  In this talk we clarify how domain
wall fermions may be used to study SYM theories \footnote{It has long been
  recognized that  SYM theory can arise
  accidentally as the low energy limit of a
  theory with gauge and chiral symmetry, and the correct fermion
  representation \cite{Kaplan:1984sk}. Lattice implementation of $N=1$ 
  SYM with Wilson fermions and fine-tuning is discussed in
  \cite{Curci:1987sm,Kirchner:1998mp,Montvay:1997ak,Montvay:1997uq}.
  Using domain wall fermions for simulation of $N=1$ SYM was suggested 
  in \cite{Nishimura:1997vg,Hotta:1997my,Hotta:1997af}, but here we follow a different
  approach. Our results parallel prior work on $N=1$ SYM theories in
  the overlap formulation \cite{Nishimura:1997hu,Neuberger:1998bg},
  which is equivalent to domain walls with
  infinite separation.}.
Throughout this talk we will actually discuss only the continuum
version, as it is simpler to formulate, if less rigorous, and there are
no technical or conceptual obstacles to translating this work
to the lattice.  We address in turn $N=1$ in $d=4$, $N=1$ in $d=3$, and 
$N=2$ in $d=4$.
\section{$N=1$ SUSY Yang-Mills theory in $d=4$ dimensions}
$N=1$ SUSY Yang-Mills theory in $d=4$ Minkowski space consists of a gauge group 
with a 
massless adjoint Majorana fermion, the gaugino.
It is expected to exhibit all sorts 
of fascinating features, such as confinement, discrete vacua, domain walls, and
excitations on these domain walls which transform as fundamentals
under the gauge group \cite{Witten:1997ep}.

The gaugino should arise as an edge
state in a 5-d theory of domain wall fermions.  The only subtlety in using the machinery of
domain wall fermions is how to obtain a single Majorana fermion in
Minkowski space, since without modification, the theory gives rise to
massless Dirac fermions in Euclidian space.

Let us first review how a massless Dirac fermion arises in the domain wall 
approach. Consider a Dirac fermion in a 5-dimensional Euclidian continuum, where
 the fifth dimension is compact: $x_5 = R \theta$, $\theta\in
(-\pi,\pi]$.  The mass of the fermion is given by a periodic step function
\beq 
m(x_5)=M\epsilon(\theta)=\cases{+M&$-\pi/2<\theta\le\pi/2$\cr
  -M&${\rm otherwise}$\cr} 
\eqn{mass}
\eeq
We introduce gauge fields independent of the coordinate $x_5$, so that 
the Euclidian action is given by
\beq 
S_5 =\int{ d^5x}\,\, i \overline\Psi D(x_5)\Psi\ ,\quad 
D(x_5)= \lb   \Dsl_4 +  \partial_5 \gamma_5 + m(x_5)\rb \ ,\qquad
\gamma_\mu=\gamma_\mu^\dagger\ .
\eqn{s5}
\eeq
Here $ \Dsl_4$ is the usual $d=4$ gauge covariant derivative for a
Dirac fermion in the adjoint representation. It is convenient to expand $\
Psi$ and $\bar \Psi$ as
\cite{Kaplan:1996pe}
\beq
\begin{array}{rcl}
\Psi(x_\mu,x_5)&=& \sum_{n} \lb{
b_n(x_5)}{P_+} +{ f_n(x_5)}
{P_-}\rb \psi_n(x_\mu)\ ,\nonumber\\
\overline\Psi(x_\mu,x_5)&=& \sum_{n}\bar \psi_n(x_\mu)\lb {b_n(x_5)}{ P_-} 
+ {f_n(x_5)}
{P_+}\rb \ .
\end{array}
\eeq
Here $P_\pm =(1\pm \gamma_5)/2$ are the chiral projection operators,
$\psi_n$ and $\bar\psi_n$ are ordinary 4-d Dirac spinors, and $b_n$,
$f_n$ form a complete basis of periodic functions satisfying the
eigenvalue equations
\beq
[\partial_5 + m(x_5)]{b_n} ={\mu_n} {f_n}\ ,\qquad [-\partial_5 + m(x_5)] {f_n} 
= 
{\mu_n} {b_n}
\ .
\eqn{susy}
\eeq
With this expansion, the action $S_5$ may be rewritten as a theory of
an infinite number of 4-d flavors with masses $\mu_n$,
\beq
S_5 = \sum_n \int { d^4x}\,\,   \bar \psi_n(x) \lb i\Dsl_4 +
i{\mu_n}\rb \psi_n(x)\ .\ 
\eqn{act4}
\eeq

 It is straight forward to solve
the above equations for $\mu_n$.  First of all, one finds zero modes 
\beq
\mu_0=0\ ,\qquad b_0(x_5) = e^{-\int^{x_5} m(y) \,dy}\ ,\qquad f_0(x_5) =
e^{+\int^{x_5}
  m(y) \,dy}\ .
\eeq
Note that $b_0$ is localized at $\theta=-\pi/2$, while $f_0$ is
localized at $\theta=+\pi/2$.  Nonzero modes 
have wave functions which are linear combinations of sines and cosines
appropriately matched at the locations of the domain walls. The
corresponding eigenvalues are doubly degenerate
\beq
 \mu_n=\sqrt{M^2 + n^2/R^2}\ ,\quad
n=\pm 1, \pm 2,\ldots
\eeq

If instead of having a kink-like mass profile for the
$\Psi$ fermions we had a constant mass $M$ (again with periodic
boundary conditions), the corresponding
eigenvalues $\bar \mu$  would be
\beq
\bar\mu_0=M\ ,\qquad \bar\mu_n=\sqrt{M^2 + n^2/R^2}\ ,\quad
n=\pm 1, \pm 2,\ldots
\eeq
Note that for $n\ne 0$, the eigenvalues $\mu_n$ and $\bar \mu_n$ are
equal.  It follows that the ratio of fermion determinants for a kink
and a constant mass is given by (assuming appropriate regularization)
\begin{figure}
\centerline{\epsfxsize=3.5in \epsfbox{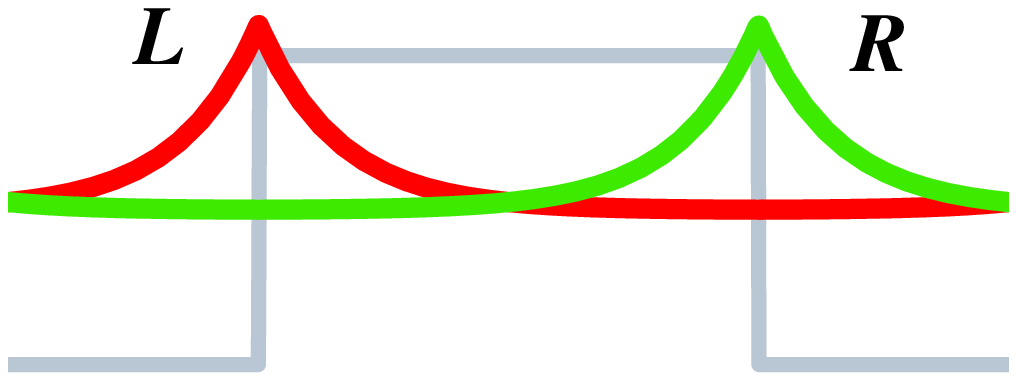}}
\begin{caption}
{{\it The left- and right-handed zeromode components, localized at the mass
kinks.} }
\end{caption}
\end{figure}
\beq
{\det\lb i(\Dsl_4 + \gamma_5\partial_5 + M\epsilon(\theta))\rb \over 
\det\lb i(\Dsl_4 + \gamma_5\partial_5 + M
)\rb } = 
{\det \lb i\Dsl_4\rb \over \det \lb  i(\Dsl_4 + M)\rb }
\eeq
Note that the right hand side of the above equation corresponds to a
massless Dirac fermion and an uninteresting Pauli-Villars field. The
left and right handed components of the massless
Dirac fermion correspond to the edge states $b_0$ and $f_0$ (see Fig.~1.). 
This method for obtaining a single massless Dirac fermion is robust when
transcribed on the lattice \cite{Kaplan:1992bt}:  the beauty of the method is that there is 
no chirality in 5-d, and if one shifts or renormalizes the fermion
mass term in the 5-d theory by $\delta m$ (with $\vert \delta m\vert < 
M$), the effective 4-d theory still has a
massless mode.  That is because there is a gap in the bulk, so that
$b_0$ and $f_0$ fall off exponentially, while a chiral symmetry
breaking fermion mass must be proportional to the (exponentially
small) overlap of $b_0$ and $f_0$.

In order to simulate $N=1$ SYM theory, we need to impose a Majorana
condition on $\psi_0$.  Note that in 4-d Minkowski space, the Majorana
condition is $\psi = C\bar\psi^T$, where $C$ is the charge
conjugation matrix, satisfying $C^{-1}\gamma^\mu C =
-{\gamma^\mu}^T$ and $C^{-1}T_a C = -T_a^T$ for generators $T_a$ of
real or pseudo-real representations of the gauge group. 
In Minkowski space charge conjugation interchanges 
left- and right-handed particles. In our Euclidian domain wall theory, 
the left- and right-handed modes live on the two different kinks.
This suggests that the correct ``Majorana'' condition for the 5-d
Euclidian theory is to define a 5-d reflection  which interchanges the 
two chiral zeromodes, $\CR_5: \theta \to
-\theta$,  and to
impose the constraint on the 5-d Dirac fermions
\beq
\Psi = \CR_5 C \bar \Psi^T
\eeq
The 5-d path integral then results in a fermion pfaffian, rather than
a fermion determinant: 
\beq
Z_5 = \pf \lb i\CR_5 C 
  \lp\Dsl_4 + \gamma_5\partial_5 + m(x_5)\rp\rb\ .
\eeq
 It is
  straightforward to check that we are taking the pfaffian of an
  antisymmetric operator, as is required \footnote{Actually, the
    operator is only antisymmetric if the fermion is in a real
    representation, such as  an adjoint, instead of a pseudoreal
    representation.  Thus our method is consistent with Witten's
    result \cite{Witten:1982fp} that  a theory of a single Weyl  pseudoreal fermion 
    is sick.}.  In terms of the
  mode expansion in 4-d fields,  note that $\CR_5$ interchanges
   $b_n(x_5) 
  \leftrightarrow f_n(x_5)$, so the constraint yields 
\beq
\begin{array}{rcl}
\Psi &=& \sum_{n}  \lb {
b_n(x_5){P_+} +}{ f_n(x_5)}
{P_-}\rb \psi_n(x_\mu) \\ &=& 
\sum_{n} \lb{
b_n(x_5)} {P_+} +{ f_n(x_5)}
{P_-}\rb C\bar \psi_n^T(x_\mu) = \CR_5 C \bar\Psi^T
\end{array}
\eeq 
which implies the conventional (Euclidian) Majorana constraint on the
4-d fermion fields:
\beq
\psi_n(x_\mu)= C\bar\psi_n^T(x_\mu)\ .
\eeq

Using the same technique as in the Dirac case to remove bulk modes,
we arrive at a formula for the pfaffian of a massless Majorana
fermion:
\beq
{\pf\lb i\CR_5 C (\Dsl_4 + \gamma_5\partial_5 + M\epsilon(\theta))\rb \over 
\pf\lb i\CR_5 C (\Dsl_4 + \gamma_5\partial_5 + M
)\rb } = 
{\pf \lb i C \Dsl_4\rb \over \pf \lb  i C (\Dsl_4 + M)\rb }
\eeq
This formula is easily extended to the lattice by replacing the Dirac
action by the Wilson action in all five dimensions
\cite{Kaplan:1993sg}. As mentioned before, this leads to an answer
identical to that derived by Neuberger \cite{Neuberger:1998bg},
although derived in a somewhat different way.  

By using
Neuberger's closed expression for the domain wall determinant, it is
possible to show that the lattice version of the above pfaffian is
positive definite, and hence can be computed unambiguously as the
square root of the Dirac determinant \footnote{This  observation was
  made to DK by Y. Kikukawa.}. Thus the domain wall
approach has an added advantage over the Wilson fermion strategy, which
suffers from a pfaffian which is not positive definite
\cite{Kirchner:1998nk}. it is therefore feasible with present
technology to begin exploring this interesting theory.

\section{$N=1$ SUSY Yang-Mills theory in $d=3$ dimensions}
$N=1$ SYM in $d=3$ is an interesting theory as well, especially in
light of Witten's recent discussion of dynamical SUSY breaking
\cite{Witten:1999ds}. Once again the spectrum consists of a gauge symmetry and a
Majorana fermion, the gaugino. There are two independent relevant
operators that break SUSY: the gaugino mass and the (quantized) Chern-Simons term, 
with one linear combination of the two being supersymmetric.  In what
follows we will assume that form some gauge groups it is possible to formulate the lattice
theory such that the  coefficient of the Chern-Simons term in
the effective 3-d continuum theory
vanishes (work in progress here!).  In that case, the only relevant
SUSY breaking operator is once again the gaugino mass.  If we can
realize chiral symmetry and gauge symmetry with a Majorana fermion,
SUSY will once again arise in the continuum as an accidental symmetry, 
modulo the unresolved issue of the Chern-Simons term.

We saw above that without constraints, a 5-d
domain wall theory led to a massless Dirac fermion in 4-d; to end up
with a Majorana fermion we had to impose a generalization of the
Majorana constraint, which effectively took the square root of the
5-d domain wall determinant.  However,
following the same procedure in one fewer dimensions, a 4-d domain
wall system with a Dirac fermion gives rise to two massless Dirac
fermions in 3-d,  four times as many degrees of freedom as we wish! In 
particular, 
\beq
 {\det\lb i (D_i\gamma_i + \gamma_4\partial_4 + M\epsilon(\theta))\rb \over 
 \det \lb i(D_i\gamma_i + \gamma_4\partial_4 + M
)\rb } = 
{  \lb\det i\Dsl_3\rb^2 \over  \lb\det  i  (\Dsl_3 + M)\rb^2 }
\eeq
where on the left hand side, the index $i$ runs from 1 to 3 and the
$\gamma$ matrices are $4\times 4$; on the right hand side, $\Dsl_3$ is the 3-d Dirac
operator ($2\times 2$ dimensional in spinor space).
Therefore it is clear we need to impose two binary constraints on the
system.

First of all, instead of using 4-d Dirac domain wall fermions, we can
impose the 4-d Euclidian Majorana constraint, $\psi = C_4 \bar \psi^T$, where $C_4$ is a 4-d charge
conjugation matrix.  This naturally gives rise to a 3-d theory with
two  Majorana fermions localized at the two kinks.  To reduce the
spectrum to a single Majorana fermion in 3-d we use the trick of the
previous section and constrain the field further to be Majorana under
the 3-d charge conjugation matrix $C_3$, and a simultaneous reflection 
$\CR_4$ in the compact fourth dimension.  Thus the simultaneous constraints are:

\begin{enumerate}
\item $\Psi(x_i, x_4) = C_4\bar\Psi^T(x_i,x_4)$,
\item $\Psi(x_i,x_4) =
    \CR_4 C_3 \bar \Psi^T(x_i,x_4)$
\end{enumerate}
To be explicit, one can choose the $\gamma$ matrix basis:
\beq
\gamma_i = \sigma_1\otimes \sigma_i\ ,\quad \gamma_4 = \sigma_3\otimes 
1\ ,\qquad
C_3 = 1\otimes \sigma_2\ ,\qquad C_4=\sigma_1\otimes \sigma_2\ .
\eeq

It isn't obvious how to simultaneously impose these two constraints
until one uses constraint (1) to replace constraint (2) by
\begin{itemize}
\item[2'.] $\Psi(x_i, x_4) =\CR_4 C_3 C_4^{-1}\Psi(x_i,x_4) $
\end{itemize}
This last constraint, relating $\Psi$ to its reflection, tells us that 
we are living on an orbifold --- only half the world we were
considering represents independent degrees of freedom.  So what we do
is impose constraint (1) and compute the path integral over half our
original space, namely for $\theta\in (0,\pi]$ with suitable
boundary conditions at the fixed points of $\CR_4$:
\beq
\lb 1- C_3 C_4^{-1} \rb\Psi(x_i,x_4)\Bigg\vert_{x_4=0,\, \pi R} = 0
\eeq
Then one finds the desired result,
\beq
{\pf\lb i C_4 (D_i\gamma_i + \gamma_4\partial_4 + M\epsilon(\theta))\rb \over 
\pf\lb i C_4 (D_i\gamma_i + \gamma_4\partial_4 + M
)\rb } = 
{ \pf \lb\, i C_3 \Dsl_3\rb \over \pf \lb \, i C_3 (\Dsl_3+ M)\rb }\ , 
\qquad \theta\in (0,\pi]\ .
\eeq
We have not yet completed our analysis of the reality/positivity of
the 4-d pfaffians on the lattice, and the related issue of the Chern-Simons term
 in the
effective 3-d theory.
\section{$N=2$ SUSY Yang-Mills theory in $d=4$ dimensions}
$N=2$ SYM in $d=4$ would be fascinating to simulate on the lattice, since in
the continuum it exhibits a vast array of interesting phenomena
\cite{Seiberg:1994rs}. One might think that it impossible to
do without fine tuning, however, because of the scalar fields in the $N=2$ gauge
multiplet.  However, a promising idea is to formulate the theory first 
as an $N=1$ SUSY theory in $d=6$ (starting from a domain wall theory
in $d=7$) \cite{Nishimura:1997hu}.  The light spectrum of the $d=6$
theory, with UV cutoff
$\Lambda_6$  would consist of
gauge fields and a Weyl fermion. Then at a scale $\Lambda_4\ll \Lambda_6$,
one compactifies to $d=4$:  the extra two gauge boson polarizations
become the complex scalar of the $d-4$, $N=2$ gauge multiplet, while
the Weyl fermion in $d=6$ becomes the required two Weyl fermions in
$d=4$.  Furthermore, all gauge, $\phi^4$ and Yukawa couplings in the $d=4$
effective theory are derived from the $d=6$ gauge coupling $g_6$.

This approach is made respectable by the fact that in the continuum,
the $N=1$ SUSY algebra in $d=6$ reduces under compactification to the
$N=2$ SUSY algebra in $d=4$ \cite{Sohnius:1985qm}.

Of course, the idea is still to have the target $N=2$ theory arise as
an {\it accidental} symmetry in the effective theory. What one must
try to do then is to take $\Lambda_4$ sufficiently smaller than
$\Lambda_6$ so that by the time one has scaled down to $\Lambda_4$ and 
passed over to the $d=4$ effective theory, the theory is ``supersymmetric enough'' 
 to ensure that the noxious scalar masses
radiatively generated in the effective $d=4$ theory are ``small
enough''.

How small is ``small enough''?  To study the $N=2$ theory in the
strongly coupled region, where it is interesting, we need the scalar
mass $m_s$ to satisfy $m_s\ll \Lambda_{SQCD}$ where $\Lambda_{SQCD}$
is the scale where the $N=2$ gauge interactions get strong.

Unfortunately this is impossible to achieve. The  $N=1$ supersymmetry
in the $d=6$ theory is only a symmetry of the operators of leading
dimension; SUSY is 
 violated by higher dimension operators, suppressed by powers of
$\Lambda_6$.  Thus the SUSY violating radiatively generated scalar
masses  in the $d=4$ effective theory will be suppressed by powers of
$\Lambda_4/\Lambda_6$.  We can suppress these terms as much as we
want, by taking this ratio to be very small!  However, the mass scale
$\Lambda_{SQCD}$ is always smaller as it is exponentially small in
$\Lambda_4/\Lambda_6$.

To understand this, define the dimensionless gauge coupling
$\hat{g}_6=g_6 \Lambda_6$ in the $d=6$ theory. Since we
begin with a weakly coupled domain wall fermion in $d=7$
$ \hat{g}_6 \lessim 1$.
The coupling of the $d=4$  theory renormalized at the compactification
scale $\Lambda_4$ is then given by
$g_4=g_6 \Lambda_4= \hat{g_6} \Lambda_4/\Lambda_6$.    
Therefore
\beq
\Lambda_{SQCD}\sim \Lambda_4 e^{-8 \pi^2/ g_4^2} \sim
\Lambda_4 e^{-8 \pi^2/\hat{g}_6^2 (\Lambda_6/\Lambda_4)^{2}}
\ll \Lambda_4 e^{- (\Lambda_6/\Lambda_4)^{2}}\ .
\eeq
We see that while we obtain scalar masses suppressed by powers of
$\Lambda_4/\Lambda_6$, the strong interaction scale $\Lambda_{SQCD}$
is {\it exponentially} suppressed in the same ratio.  It follows that
one cannot study the $N=2$ theory in the interesting strongly
interacting regime starting
from a weakly coupled domain wall in $d=7$, without fine tuning.  

The
above argument does not rule out studying $N=2$ SYM in $d=3$ by
compactifying a $d=4$ theory with approximate $N=1$ supersymmetry,
since the gauge coupling in $d=3$ does not run logarithmically.
However, this $d=3$ theory has no ground state in the continuum, and
so it does not seem interesting to simulate.
\section{Conclusions}
Domain wall fermions offer a compelling advantage over Wilson fermions 
in simulating $ N=1$ supersymmetric Yang-Mills theories on the lattice 
in $d=4$ and $d=3$. In each case, supersymmetry arises as an
accidental symmetry, without fine-tuning.
Both of these theories should be interesting to study in the near future. 

As for SUSY theories with scalars: it is hard to imagine how one can
evade fine-tuning --- after all, if one did have such a method, it would 
provide an alternative to SUSY as a solution to the hierarchy problem!

It would be interesting to study perfect supersymmetric  actions
to try to extract the analogue of a Ginsparg-Wilson relation for
supersymmetry, for then one might identify a clever approach to SUSY
theories with scalars in the spectrum, one that minimizes the
fine-tuning problems.

%
%
\bibliography{susyym}
\bibliographystyle{h-elsevier2.bst}

\end{document}